\documentclass[prl,twocolumn,showpacs,preprintnumbers,amsmath,amssymb,superscriptaddress]{revtex4}

\usepackage{graphicx}
\usepackage{dcolumn}
\usepackage{bm}

\begin{document}

\title{Enhanced sensitivity to variation of $m_{e}/m_{p}$ in molecular spectra}

\author{D. DeMille}
\affiliation{Department of Physics, Yale University, New Haven, CT 06520, USA}

\author{S. Sainis}
\affiliation{Department of Physics, Yale University, New Haven, CT 06520, USA}
\altaffiliation{present address: Dept. of Mechanical Engineering, Yale Univ., New Haven, CT 06520 USA}

\author{J. Sage}
\affiliation{Department of Physics, Yale University, New Haven, CT 06520, USA}
\altaffiliation{present address: MIT Lincoln Laboratory, Lexington, MA 02420 USA}

\author{T. Bergeman}
\affiliation{Department of Physics and Astronomy, SUNY, Stony Brook,
NY 11794, USA}

\author{S. Kotochigova}
\affiliation{Physics Department, Temple University, Philadelphia, PA 19122, USA}

\author{E. Tiesinga}
\affiliation{Joint Quantum Institute and Atomic Physics Division, National Institute of
Standards and Technology, Gaithersburg, MD 20899}

\date{\today}

\begin{abstract}
We propose new experiments with high sensitivity to a possible spatial or temporal variation of the electron-to-proton mass ratio $\mu \equiv m_e/m_p$.  We consider a nearly-degenerate pair of molecular vibrational levels, where each state is associated with a different electronic potential.  The change in the splitting between such levels, with respect to a change in $\mu$, can be large both on an absolute scale and relative to the splitting.  We demonstrate the existence of such pairs of levels in Cs$_2$.  The narrow spectral lines achievable with ultracold Cs$_2$ in these long-lived levels make this system promising for future searches for small variations in $\mu$.
\end{abstract}

\pacs{06.20.Jr,33.80.Ps,33.20.Wr,33.20.Bx}
\maketitle

The notion that the fundamental constants of nature may actually vary
has recently generated intense interest \cite{Murphy01,Uzan03}.  Theoretical motivation comes from a variety of string-inspired models; these can include space-time with extra dimensions of variable geometry, and/or light scalar fields whose variable amplitude couples to ordinary matter. Both effects can change the apparent values of constants.  Such fields are potential candidates (dubbed ``quintessence'') to explain the observed dark energy that dominates the universe.  Sensitive probes for possible variation of fundamental constants are among the few ways to verify or constrain models such as these.

On quite general principles, measurements can only detect variation in \textit{dimensionless} constants \cite{Uzan03}.  To date,  most attention has focused on possible variation of the fine structure constant $\alpha$.  However, variations in $\alpha$ are no more likely than in any other fundamental dimensionless parameter.  In fact, it has been argued that within the context of grand unified theories, variation in the electron-to-proton mass ratio $\mu \equiv m_e/m_p$ should be larger than the variation in $\alpha$ by a factor of $\sim \! 30$ \cite{Uzan03}.  As a result, study of $\mu$ is a potentially more sensitive method to observe any variation of constants. 

The most sensitive tests for variation in
$\mu$ have come from examining molecular spectra from
cosmologically distant sources, at redshifts corresponding to $\sim\!10^{10}$ years ago \cite{cosmodmu}.  Comparisons of spectral lines in these objects to the same lines
measured now in the laboratory, yield stringent bounds at the level
$\sim\! 10^{-15}$/year on the time variation of
${\mu}$. However, systematic errors associated with these measurements can be significant, and are difficult to eliminate due to the
uncertainty in the structure, environment, and dynamics of the sources.

We propose a new type of laboratory experiment to search
for variation of $\mu$ with high sensitivity.  We
show that the energy of molecular vibrational levels can be highly sensitive to changes in $\mu$.  In addition, pairs of closely-spaced levels can be used to enhance the sensitivity relative to the level splitting.  We show spectroscopic data verifying such a near-degeneracy in the ground vibronic levels of Cs$_2$.  We argue that the splitting between pairs of levels in Cs$_2$ could be measured precisely enough to sense a fractional change $\Delta \mu/\mu \lesssim 10^{-17}$.

The sensitivity of our proposed measurement is governed by a few key parameters. 
Most important is the absolute change $\partial_{\mu} \Omega$ of the 
energy splitting $\Omega$ with respect to a fractional change $\Delta \mu / \mu$: $\partial_{\mu} \Omega \equiv \partial \Omega / \partial (\mathrm{ln} \mu)$.
($\hbar = c =1$ throughout.)  $\Omega$ can be measured with statistical uncertainty $\delta \Omega = \Gamma/S$, where $\Gamma$ is the linewidth of the transition and $S$ is the signal-to-noise.  Hence, $\Delta \mu / \mu$ can be detected with statistical uncertainty $\delta (\Delta \mu / \mu) = \delta \Omega /\partial_{\mu} \Omega = (\Gamma / \partial_{\mu} \Omega)S^{-1}$, and the ratio $\partial_{\mu} \Omega/\Gamma$ provides a primary figure of merit for any such measurement.  However, the \textit{relative} change of the splitting, $\partial_{\mu} \Omega/\Omega$, is also of importance, for two reasons.  First, since ultimately $\Omega$ must be measured with respect to some reference clock (with fractional uncertainty $\delta \Omega_c/\Omega_c$), it is impossible to determine $\Omega$ to better than $\delta \Omega_{\mathrm{min}} = \Omega (\delta \Omega _c/\Omega _c)$.  In addition, many systematic effects are proportional to $\Omega$ (e.g. Doppler shifts).  Hence, $\partial_{\mu} \Omega/\Omega$ provides an important secondary figure of merit.

We analyze the shifts in molecular energy levels when $\mu$ changes, with the assumption that both $m_e$ and $\alpha$ remain constant.  In this case, a change in $\mu$ corresponds to a variation of $m_p$; in addition, molecular electronic potentials are independent of $\mu$, since such potentials depend parametrically only on the Rydberg energy $Ry = m_e \alpha^2/2$.  (The assumption of constant $\alpha$ and $m_e$ is equivalent to using a reference clock based on optical atomic transitions, which also depend parametrically only on $Ry$ and $\alpha$.)  We approximate the vibrational energy levels $E_v$ via the WKB quantization condition:
\begin{equation} \label{WKB}
\int _{R_i} ^{R_o} \sqrt{2 M [E_v - V(R)]}dR = (v+ \frac{1}{2})\pi,
\end{equation}
where $R$ is the internuclear distance, $V(R)$ is the potential with minimum value 0, $R_i (R_o)$ is the classical inner (outer) turning point of $V(R)$ at energy $E_v$, and $M \propto m_p$ is the reduced mass of the nuclei in the molecule.  Varying Eq. (\ref{WKB}) with respect to both $\mu$ and $v$, one obtains for the \textit{energy sensitivity} $\partial_{\mu} E_v$:
\begin{equation}\label{dE}
\partial_{\mu} E_v \equiv \frac{\partial E_v}{\partial (\mathrm{ln} \mu)} = \frac{(v+ \frac{1}{2})}{2 \rho (E_v)},
\end{equation}
where $\rho(E_v) = (\partial E_v/\partial v)^{-1} \approx (E_{v}-E_{v-1})^{-1}$ is the density of states at energy $E_v$ \cite{Chin06}.  Near its minimum, a typical potential is harmonic, with $E_v = (v+\frac{1}{2})\omega$ and $\rho(E_v) = \omega^{-1}$ ($\omega$ is the classical oscillator frequency). Hence, for the harmonic part of the potential, $\partial_{\mu} E_v = (v+ \frac{1}{2})\omega/2 = E_v/2$ has a $\sim\! v$-fold enhancement for the $v^{th}$ vibrational level. (This can also be derived from $\omega \propto M^{-1/2}$.)  This constitutes a general mechanism for amplifying the \textit{absolute} size of $\partial_{\mu} \Omega$.  

For sufficiently large values of $v$, real molecular potentials are not harmonic.  For highest values of $v$, $E_v$ approaches the dissociation limit $D$, and the density of states $\rho (E_v)$ becomes large.  Hence, the value of $\partial_{\mu} E_v$ again becomes small for the highest levels in any potential.  Therefore, at some intermediate value of $v$, $\partial_{\mu} E_v$ is maximized.  We verified this behavior both for a generic (Morse) potential with analytic expressions for $E_v$, and for the experimentally determined $X ^1\Sigma_g^+$ potential of Cs$_2$ (see below).  In both cases, the maximum value of $\partial_{\mu} E_v$ is obtained for levels with $E_v^{\mathrm{(max)}} \approx 3D/4$, and there the energy sensitivity is only slightly diminished from its expected value in the harmonic approximation: $\partial_{\mu} E_v^{\mathrm{(max)}} \approx (2/3)(E_v^{\mathrm{(max)}}/2)$.  Note that a recent paper \cite{Chin06} discussed the propect for detecting variations in $\mu$ using Feschbach resonances in the scattering between ultracold atoms.  However, such resonances arise from the presence of vibrational states (associated with potentials for other internal states of the atoms) that are very near $D$, where $\partial_{\mu} E_v$ is small.  The scheme discussed here uses the much larger energy sensitivity for states of intermediate vibrational excitation.

The same mechanism can be used, in some cases, to provide transitions with extremely large \textit{relative} shifts $\partial_{\mu} \Omega/\Omega$.  Consider a situation in which two molecular electronic potentials $X$ and $Y$ overlap, but the minimum of potential $Y$ (at energy $T_Y$) is at higher energy than that of $X$ (at $0$).  In this case, an excited level of $X$ with vibrational number $v_X \gg 1$ and energy $E_1 = E^{(X)}_{v_X}$ can be quite near in energy to a lower vibrational level of $Y$ ($v_Y \ll v_X$), with energy $E_2 = T_Y + E^{(Y)}_{v_Y}$. The energy difference between such a pair of levels is small ($\Omega = |E_1 - E_2| \ll E^{(X)}_{v_X}$), but can retain a large sensitvity to changes in $\mu$, since $\partial_{\mu} \Omega = \partial_{\mu}E^{(X)}_{v_X} - \partial_{\mu}E^{(Y)}_{v_Y} \approx \partial_{\mu}E^{(X)}_{v_X}$.

The molecule Cs$_2$ is an attractive system for implementing this scheme.  Cs$_2$ has two low-lying, overlapping potentials: the deep $X ^1\Sigma_g ^+$ ground state, and the shallower $a ^3\Sigma_u ^+$ state, each of which dissociates to a pair of ground-state Cs atoms \cite{Herzberg79}.  The vibrational splittings of these two potentials are incommensurate, and the large mass of Cs yields a high density of states; hence near-degeneracies appeared likely.  In fact, Ref. \cite{Weickenmeier86} reported evidence for such in the high-$J$ levels of the $v_X = 137$ state.  Furthermore, all rovibrational levels of both potentials have extremely long radiative lifetimes ($\gg 1$ s) \cite{Vanhaecke02}, and production of \textit{ultracold} Cs$_2$ molecules (which can be measured over long coherence times) has become routine \cite{Vanhaecke02,ultracoldCs2}; hence the system can yield spectral lines with narrow width $\Gamma$. Finally, it has been predicted that deeply-bound levels of the $a ^3\Sigma_u ^+$ state--which lie near $E_v^{\mathrm{(max)}}$ for the $X ^1\Sigma_g ^+$ state--can be efficiently populated \cite{Drag00,Vatasescu00}.

\begin{figure}
\centering
\includegraphics[width=3.4in]{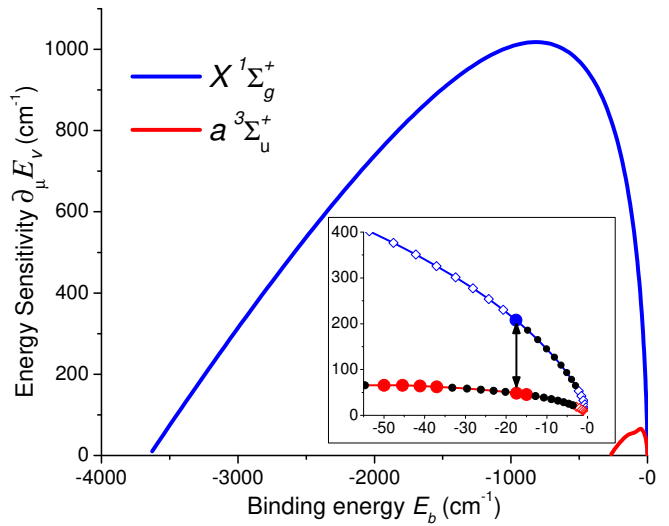}
\caption{Energy sensitivity vs. binding energy, for the $X ^1\Sigma_g^+$ and $a ^3\Sigma_u^+$ states of Cs$_2$.  Inset: a zoom into the region discussed here.  Colored circles are levels observed in this work; open diamonds are levels from previous data; small black dots are predictions from our fitted potentials. The arrow indicates the degeneracy described in the text.} \label{fig01}
\end{figure}

We explicitly calculate $\partial_{\mu} E_v$ for all vibrational levels in the $X$ and $a$ states of Cs$_2$, using the fitted potentials described below and numerical solutions for $E_v$ with slightly different values of $M$.  (Similar results were obtained very simply for all but the highest-$v$ levels of the $X$ state, using a Dunham expansion for $E_v$ \cite{Weickenmeier86,Herzberg79}.)  A plot of $\partial_{\mu} E_v$ versus binding energy ($E_b(v)\equiv E_v - D$) is shown for both potentials in
Fig. \ref{fig01}. The behavior throughout is as expected from Eq. \ref{dE}.  Note that indeed, as expected from the previous discussion, $\partial_{\mu} E^{(a)}_{v_a} \ll \partial_{\mu} E^{(X)}_{v_X}$ for close-lying $X$ and $a$ state levels (such that $E_b^{(a)}(v_a) \approx E_b^{(X)}(v_X)$).  In addition, $\partial_{\mu} E^{(X)}_{v_X}$ increases monotonically with $|E_b|$, throughout the region of $a$ state levels.  Hence, a search for variation of $\mu$ will be most sensitive when using nearly-degenerate levels lying as close as possible to the $a$ state minimum.

We experimentally determine the positions of deeply bound $a ^3\Sigma_u^+$
levels, using two-color photoassociation (PA) spectroscopy \cite{Abraham95,Vanhaecke04}.   A sample of ultracold ($T\!\sim 100\,\,\mu$K) Cs atoms is cooled and confined in a dark SPOT magneto-optic trap.  After applying a uniform magnetic field, the atoms are optically pumped into the $6s_{1/2}\, F\! =\! m_F\! =\! 4$ state.  Next a
CW laser (the PA laser) excites the atoms into a bound state \cite{Fioretti99} of the Cs$_2$ $(2) 0_g^-$ potential (dissociating to the $6s_{1/2}+6p_{3/2}$ asymptote). This excited state decays to a manifold of
$a ^3\Sigma_u^+$ vibrational levels, which are ionized by a pulsed laser.
A second CW (probe) laser is applied to search for resonances between the desired $a ^3\Sigma_u^+$ levels and the $(2) 0_g^-$ level excited by the PA laser. On such a resonance, the probe beam
can shift the excited state off resonance with the PA
laser and decrease the rate of molecule formation.  We scan the
probe laser frequency while monitoring the ion yield, and look
for resonant dips in the signal.  The difference between the frequencies of the resonant probe and PA lasers gives the binding energy of the $a ^3\Sigma_u^+$ level.  We report results for $E_b$ relative to the hyperfine (HF) barycenter, located $\approx 8.04$ GHz below the $6s_{1/2}F\! =\! 4 + 6s_{1/2}F\! =\! 4$ asymptote.  Use of PA levels in the pure long-range well of the $(2)0_g^-$ state
enables access to deeply bound $a ^3\Sigma_u^+$ levels, because of favorable Franck-Condon factors \cite{Drag00,Vatasescu00}.  

Each ground-state vibrational level has considerable
substructure.  We write $a$ and $X$ states in the basis $|(S,I)f,\ell,\mathcal{F}\rangle$, where $I$ is the total spin of the pair of Cs nuclei; $S$ is the total electron spin; $\mathbf{f}\equiv\mathbf{S}+\mathbf{I}$; $\ell$ is the rotational angular momentum; and $\mbox{\boldmath{$\mathcal{F}$}}\equiv\mathbf{f}+\mbox{\boldmath{$\ell$}}$.  For the $a ^3\Sigma_u^+$ states of interest, $S=1$ and the HF interaction dominates the sublevel structure, spliting levels with different $f$ according to $E_{HF} = \frac{A}{4}[f(f+1)-S(S+1)-I(I+1)]$, where $A \approx 2.3$ GHz is the Cs $6s_{1/2}$ HF constant.  For each value of $f$, there is a manifold of closely-spaced states: those with different $\ell$ are split by the rotational energy $B_v\ell(\ell+1)$ ($B_v \sim 0.1$ GHz), and states with different (same) $\mathcal{F}$ are split (mixed) by a $2^{\mathrm{nd}}$-order spin-orbit (SO2) interaction of comparable strength \cite{Kotochigova00}.  For most states we consider, the eigenstate is well-approximated by a basis state.  However, levels of the $X ^1\Sigma_g ^+$ state ($S=0$) can mix with nearby $a^3 \Sigma_u ^+$ levels via HF interactions.  


Pairs of Cs atoms in the spin-polarized sample have $f\! =\! m_f\! =\! 8$, and only pairs with $\ell\! =\! 0$ ($s$-wave scattering state) are excited by the PA laser; hence initially $\mathcal{F}\! =\! m_\mathcal{F}\! =\! 8$.
The Hund's case (c) $(2)0_g^-$ excited state has resolved levels of definite $J'$, where $\boldmath{J}'\! =\! \mbox{\boldmath{$\ell$}}'\! +\! \mathbf{S}'\! +\! \mathbf{L}'\!$, $L'\! =\! 1$ is the electron orbital ang. mom., and $S'\! =\! 1$. The PA laser is tuned to excite only $J'\! =\! 2$ levels, which are a mixture of $\ell'\! =\! 0,2,4$ \cite{Jones96}.  The PA laser has $\sigma^+$ polarization, so electric dipole (E1) selection rules ensure $\mathcal{F}'\! =\! M_{\mathcal{F}}'\!= \! 9$. Since $\mbox{\boldmath{$\mathcal{F}$}}'={\bf J}'+{\bf I}'$ and $\mathcal{F}'=9$, only $I'=7$ is excited.  E1 selection rules for the probe transition ensure that only a$^3\Sigma_u^+$ states with $I\! =\! 7$ and $\ell\! =\! 0,2,4$ are detected; with polarization $\sigma^-$ ($\sigma^+$), only sublevels with $\mathcal{F}\! =\! M_{\mathcal{F}}\! =\! 10$ ($\mathcal{F}\! =\! 8,9,10$, $M_{\mathcal{F}}\! =\! 8$) are seen.

We observe multiple sublevels for several $a^3 \Sigma_u ^+$
vibrational levels with $|E_b(v)|\! \sim\! 400\! -\! 1500$ GHz.  The level positions and line strengths qualitatively agree with predictions based on the $a^3 \Sigma_u ^+$ long-range potential \cite{Vanhaecke04}, which yielded approximate values for $E_b(v)$ and $B_v$.  These predictions guided an initial assignment of quantum
numbers to the observed states. As expected, sublevels for each $v$ are clustered together according to their $f$ value, and for each $f$ we typically observe several sublevels with different values of $\ell$ and $F$.  We perform a global fit of the potentials to all known levels \cite{Weickenmeier85,Vanhaecke04} in both the $a ^3\Sigma_u ^+$ and $X ^1\Sigma_g^+$ states (including our data), taking into account the SO2 interaction.  For spectral regions with no or insufficient data, the potentials were modeled using the ``Hannover'' analytic form \cite{Samuels00}.  The quality of the fit to this data suggests that we have a good understanding of the level assignments and interaction strengths: the r.m.s. deviation was $\approx 90$ MHz for HF/rotational/SO2-induced sublevel splittings, and $\approx 200$ MHz for absolute binding energies, vs. typical experimental uncertainties of $\approx 30$ MHz  and $\approx 400$ MHz, respectively.  Details of the data and the fit will be given elsewhere.  For now, we focus on a specific feature in the data.  

\begin{figure}
\centering
\includegraphics[width=3.4in]{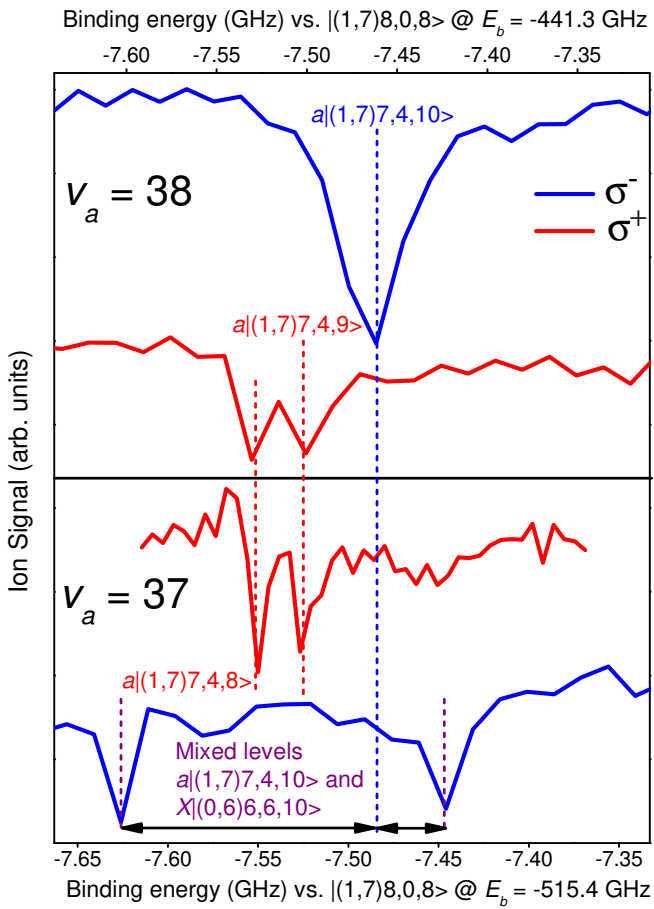}
\caption{Data indicating a degeneracy between the deeply bound $a ^3\Sigma_u^+(v_a=37)$ and $X ^1\Sigma_g^+(v_X=138)$ levels in Cs$_2$.  Ion signals are plotted vs. probe laser frequency (relative to the line from the $a |(1,7)8,0^+,8\rangle$ state in the same vibrational level), for two adjacent vibrational levels tentatively assigned as $v_a=38,37$ (upper and lower panels, respectively).  The label $\sigma^{\pm}$ refers to the polarization of the probe laser.  Frequency scales are slightly offset (by $\sim\! 20$ MHz) to align the $\mathcal{F}=8,9$ spectral lines, and hence to emphasize the similarity between the two $\sigma^+$ spectra.  The dashed vertical lines are meant to guide the eye in comparing the relative line positions.}\label{fig02}
\end{figure}

Fig. \ref{fig02} shows data corresponding to levels in the $f = 7$ manifold of states for two adjacent vibrational levels of the $a^3 \Sigma_u ^+$ state, with $|E_b(v)| \approx 523$ and $449$ GHz, tentatively assigned from our fits as $v_a=37$ and $38$ respectively.  Since only a single $\mathcal{F}=10$ level in the $f=7$ manifold satisfies all selection rules, we expect only a single line with $\sigma^-$ probe polarization.  This is indeed observed for the $v_a=38$ state, and we hence assign this level as $|(1,7)7,4,10\rangle$.  In the same spectral region, using a $\sigma^+$ probe, we observe two nearby lines of comparable strength that we assign to states $|(1,7)7,4,8\rangle$ and $|(1,7)7,4,9\rangle$.  In the $v_a=37$ level, and again with a $\sigma^+$ probe, we observe a very similar spectrum: two lines of similar strength, with a similar splitting, and separated by nearly the same amount from sublevels with $f=6,8$ in the same vibrational level.  However, in this $v_a=37$ level, with the $\sigma^-$ probe we observe \textit{two} lines in the $f=7$ manifold, each of similar strength but roughly two times weaker than the single analogous line in the $v_a=38$ level.  The only plausible mechanism to explain this additional line is that the single accessible $a ^3\Sigma_u^+$ level $|(1,7)7,4,10\rangle$ has mixed strongly with a nearby $X ^1\Sigma_g^+$ level \textit{only in this vibrational state}, due to the combined HF and SO2 interactions.  Based on our global fits, we assign this perturbing level as the $X ^1\Sigma_g^+(v_X=138)$ state $|(0,6)6,6,10\rangle$.

A sensitive search for variation of $\mu$ can be accomplished by measuring the small energy splitting between any of several sublevels in the $v_a=37/v_X=138$ manifold.  Ultracold Cs$_2$ molecules can be produced in any of the $a ^3\Sigma_u^+$ levels observed here, either by PA or via Feshbach resonance \cite{ultracoldCs2} followed by stimulated pumping to the desired level \cite{Winkler07,Sage05}.  By launching the molecules (or precursor atoms) in a manner like that used for atomic fountain clocks, a linewidth $\Gamma\!\sim 1$ Hz could be achieved \cite{Marion03}.  A microwave field could drive transitions between sublevels of the $a ^3\Sigma_u^+(v_a\! =\! 37)$ and $X ^1\Sigma_g^+(v_X\! =\! 138)$ states (despite the fact that these transitions are nominally forbidden).  The strong HF mixing reduces the value of $\partial_{\mu} E_v$ for the directly observed $X|(0,6)6,6,10\rangle$ sublevel; however, from our data and fits we can accurately predict the position of other nearby, unobserved $X$ state sublevels for use in the proposed experiment.  For example, a magnetic dipole transition can be driven between the observed level $a|(1,7)6,4,10\rangle$ and $X|(0,6)6,4,10\rangle$, at $\Omega \approx 6.3$ GHz, with transition amplitude $M \approx 0.1\,\,\mu_B$ induced by HF mixing of $a$ and $X$ (here $\mu_B$ is the Bohr magneton). Alternatively, an E1 transition can be driven between the observed level $a|(1,7)8,2,10\rangle$ and $X|(0,7)7,3,10\rangle$ at $\Omega \approx 9.8$ GHz, with $M \approx 1.5\times 10^{-4} ea_0 \approx 0.04\,\,\mu_B$ induced by spin-orbit effects (here $e$ is the electron charge and $a_0$ is the Bohr radius).  

To estimate the absolute sensitivity, we conservatively assume one measurement every 2 sec with $S=100$ (vs. $S\!\sim\! 1000$ every $\sim\! 1.3$ sec for the atomic fountains in \cite{Marion03}), yielding $\delta \Omega \sim\! 5 \times 10^{-5}$ Hz in one day of integration.  From Fig. \ref{fig01}, $\partial_{\mu} \Omega \sim 5 \times 10^{12}$ Hz for transitions of this type. Hence, $\delta(\Delta \mu/\mu) \sim 10^{-17}$ could be achieved in $\sim\! 1$ day.  This requires only $\delta \Omega / \Omega \sim 10^{-14}$, i.e. over ten times less accuracy than the best atomic clocks (see e.g. \cite{Marion03}). Our fits also predict near-degeneracies at energies where $\partial_{\mu} E_v$ is considerably larger, but these await further data for confirmation. 
Note: during preparation of this manuscript, we learned of two papers exploring similar ideas in other systems \cite{Flambaum07b}. 

This work was supported by DOE, ONR, ARO, and NSF grants DMR0325580 and PHY0354211. We thank P. Hamilton and S.B. Cahn for experimental assistance and E.R. Hudson for helpful comments.                                          


\end{document}